\documentclass[12pt]{article}
\usepackage{epsfig}
\begin{document}
%set bwid 6; set fwid 6; set hwid 6; set pwid 6; set xwid 8; set ywid 6;
%igset lwid 6

\begin {center}
{\Large \bf An alternative explanation for the dibaryon suggested by experiments
at the WASA facility at Julich.}
\vskip 5mm
{D V Bugg \footnote{email: david.bugg@stfc.ac.uk}\\
{\normalsize  \it Queen Mary, University of London, London E1\,4NS, UK} \\
[3mm]}
\end {center}

\begin{abstract}
%\noindent
A series of publications of the WASA collaboration culminates in a recent paper of
Pricking, Bashkanov and Clement \cite {Pricking} claiming a $\Delta \Delta$ dibaryon
resonance at 2370 MeV.
However, as explained here, there are logical flaws in this result. 
A natural alternative arises from the reaction $pd \to NN^* (1440)p_s$, where
$p_s$ is a spectator proton.
There is supporting evidence from a recent experiment of Mielke et al. on 
$dp \to \, ^3\, He \, \pi^+ \pi^-$.
\end{abstract}

There has been a series of publications by the WASA collaboration working first at
CELSIUS, then at Julich, over the years 1997 to the present. 
The most relevant will be referenced here.
They have located a signal at 2370 MeV which they claim to be a ``hidden'' dibaryon.
They have studied many reactions using a variety of kinematics to locate this signal.
The dibaryon is claimed to exist in pd scattering to 5-body final states:
$pd \to \Delta \Delta p_s$, where $\Delta$ refers to $\Delta (1232)$.
At 2370 MeV. the average $\Delta$ mass is 1185 MeV. 
At this mass, the average phase of each $\Delta$ is just below $45^\circ$.
Why should two subliminal $\Delta$ produce a dibaryon resonance?

In Ref. [1], WASA claim a 0.8 mb $I=0$ signal in the final state $np\pi^+\pi^-$. 
This is a factor $\sim 14$ smaller than the peak cross section
observed in $I=0$ total cross sections at Rutherford Lab \cite {RAL}.
These total cross sections had a statistical accuracy of $\pm 0.1\%$ and point to
point errors of $\pm 0.2\%$.
There was a possible overall normalisation error of 2 mb but varying slowly with
momenta over the entire range up to 8 GeV/c; the systematic uncertainty
over the $\Delta \Delta$ peak is $<0.4$ mb.
There was no sign of a dibaryon in total cross sections near the $\Delta \Delta$ peak.
One would expect such a resonance to lie close to the peak of the $\Delta \Delta$ 
cross section, but there is no sign of any such effect in the total cross sections 
of Ref. \cite {RAL}.
 
\section {An alternative explanation}
There is an obvious alternative explanation for the peak at 2370 MeV from the 
production of the final state $NN^*(1440)$, whose masses add to 2379 MeV, well within
errors of the $N^*(1440)$ mass.
Fig. 1 sketches the kinematics of this process, which generates 5 particles in the final
state from 3 in the initial state.
Subscripts f and s in the figure identify fast and slow particles in the lab frame.
The top of the figure displays the production of a neutral $N^*(1440)$ and the lower
part shows production of a charged $N^*(1440)$. 
Interchanging the $\pi ^-$ and $\pi ^+$ in the final state generates $\Delta ^-$ and
$\Delta^0$ intermediate states. 
So there are 4 configurations of $\Delta$ whose Clebsch-Gordan
coefficients are taken into account in fitting experimental data.
The $\Delta (1232)\pi$ branching ratio of $N^*(1440)$ is $20-30\%$ according to
Particle Data Tables \cite {PDG}.
%Fig. 1
\begin{figure}[htb]
\begin{center}
\vskip -30mm
\epsfig{file=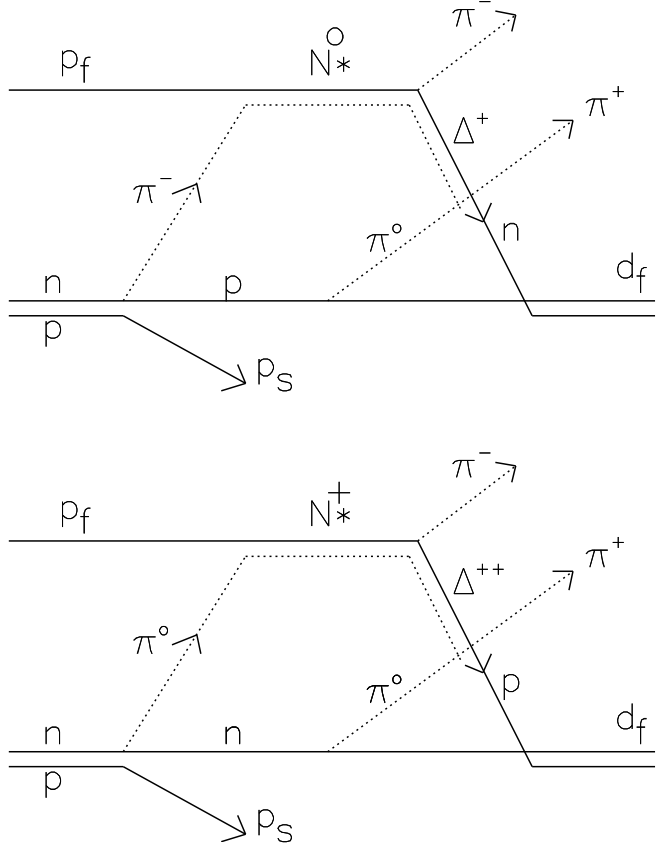,width=15cm}
\vskip -15mm
\caption {The kinematics of the reaction $pd \to NN^* \to \pi ^+\pi^-dp_s$ }
\end{center}
\end{figure}

A prominent effect in WASA data is the ABC effect of Abashian, Booth
and Crowe \cite {Abashian} \cite {Booth}.
F\" aldt and Wilkin first outlined this effect  \cite {Wilkin}; 
Gardestig, F\" aldt and Wilkin {\cite G} extended it to the reaction $pd \to \pi\, ^3He$ 
to explain an early result from WASA at CELSIUS. 
Production of $^3 He$ is favoured if the pion has a low momentum; 
otherwise production of a low momentum pion would be 
strongly suppressed by Chiral Symmetry Breaking.
The end result is that there are 3 coupled channels: $^3He\, \pi$ from the ABC
effect, together with production of $d\pi ^+$ and $d\pi^-$ pairs in Fig. 1.

The recent paper of Mielke et al. \cite {Mielke} presents new data using the
inverse kinematics to the COSY-WASA experiment: $dp \to \, ^3He\, \pi^+\pi^-$ rather 
then $pd \to \Delta \Delta \pi$.
They detect the well known ABC enhancement in the $^3He\, \pi ^+$ and $^3He \, \pi^-$
spectra, but differences between data for these two channels require some
interfering $\pi \pi$ production at low excess energies.
They model these differences in terms of $N^*(1440) \to [\Delta (1232)\pi]_{L=1}
\to N\pi \pi$, where $L$ is the orbital angular momentum between $\Delta$ and $\pi$.

Readers are referred to a paper of Bashkanov {\it et al} on $pp \to nn\pi^+\pi^-$ at 
1.1 GeV \cite {Bashkanov}. 
These data reveal a definite peak in the fourth panel of their Fig. 3 at $\sim 1400$
MeV, appearing slightly lower than 1440 MeV because of limited phase space. 
One would expect a similar peak in $np \to NN^*(1440)$. 
WASA instead attributed the peak to $\Delta (1600)$.
However, $\Delta \pi$ decays of $S_{11}(1640)$ have orbital angular
momentum $L=2$.
The centrifugal barrier factor already suppresses this decay mode near threshold.
In fact, this $L=2$ barrier distinguishes $S_{11}(1640)$ from $S_{11}(1535)$
distinctively.  
So its low momentum tail cannot be as narrow as the peak seen in WASA data.

Why is the peak at 2370 MeV so narrow?
WASA find a width of 70 MeV. 
This is much narrower than the width quoted for $N^*(1440)$ in Particle
Data tables \cite {PDG}.
However, there is a straightforward origin of this effect.
Readers are referred to a paper of Nakamura \cite {Nakamura} which provides a figure 
showing the variation of the real and imaginary parts of the $N^*(1440)$ amplitude.
The real part of the amplitude peaks at 1370 MeV and the imaginary
part peaks at 1460 MeV. 
Accordingly, there is a $45^\circ$ change in the phase of $N^*(1440)$
over this mass range.
This phase variation will introduce structure into the interference between
$NN^*(1440)$ and $\Delta \Delta$ over a narrow range of 90 MeV using
Nakamura's estimate of the mass, or 70 MeV using the PDG mass. 
This phase variation cannot be calculated a priori because of the three isobar model phases
for the three coupled channels from the ABC effect and production of 
$d\pi ^+$ and $d\pi ^-$.
These phases arise from multiple scattering amongst the 5 particles in the 
final state.
It is therefore necessary to fit phases to the WASA data.
 
An early paper of Bashkanov {\it et al} shows in their Fig. 4 a blue curve for
$pn \to d\pi ^0\pi ^0$ with a long tail extending to $>3$ GeV \cite {Bash2}. 
This tail can arise from interference between the $NN^*(1440)$ signal 
and the remains of a $\Delta \Delta$ signal. 
Their red curve shows the cross section for $pn \to d\pi ^+\pi ^-$.
A later paper of Adlarson {\it et al} \cite {Adlarson} revises the result
for $pn \to d\pi ^+\pi^-$. 
In Fig. 1 of this paper, the peak at 2370 MeV of  $pn \to d\pi ^0\pi ^0$ moves
down slightly and a rising tail appears, joining on to the earlier results
of Bashkanov {\it et al}. 
The difference between $d\pi ^0\pi ^0$ and $d\pi^+\pi^-$ can arise from different 
isobar model phases for the three coupled channels $^3 He \, \pi$ of the ABC effect 
and production of $d\pi ^0$ or $d\pi ^+ + d\pi ^-$.
These effects are capable of explaining the 0.8 mb cross section evaluated
for the peak at 2370 MeV in the latest paper of Pricking {\it et al} \cite {Pricking}. 

Consider next decays of $N^*(1440)$. 
Below a mass of 1204 MeV, phase shifts of Carter et al \cite {Carter} are 
negative; this arises from the level repulsion between the nucleon and 
$N^*(1440)$ which have the same quantum numbers. 
The phase shift then rises rapidly and reaches $20^\circ$  at 1320
MeV and the analysis of Nakamura shows that it reaches $45^\circ$
at 1370 MeV; over this mass range, decays to $N\sigma$ are dominant.
Then decays to $\Delta (1232)\pi$ rise rapidly as its phase space
increases. 
This channel accounts for the rapid rise of the imaginary part of the
$N^*(1440)$ amplitude.

There is a corollary concerning the figure of Nakamura \cite {Nakamura}.
The decay of $N^*(1440) \to \Delta (1232)\pi$ produces a P-wave pion.
The real and imaginary parts of amplitudes shown in his figure resemble a
broad P-wave cusp in the $\Delta (1232)\pi$ channel.
The real part of the amplitude peaks first, followed by a peak in the
imaginary part of the amplitude; this is characteristic of a
cusp in the P-wave, where the centrifugal barrier delays the peak in the
imaginary part \cite {Bugg}.  
                                                                         
In the late 1970's there were claims for the existence of several dibaryons 
in the reaction $pp \to d\pi^+$ over lab kinetic energies of 600--800 MeV 
in $^1D_2$, $^3F_3$ and  $^3P_2$ partial waves.
However, these claims were eventually disproved by a partial wave
analysis which included many sets of spin dependent data produced by
the Geneva group working at PSI \cite {Geneva} and by several groups
at LAMPF in the range of lab kinetic energies 500-800 MeV.     
In addition, many spin dependent measurements were made at LAMPF
using a polarised proton beam whose spin could be rotated to three
orientations: the N direction, normal to the plane of scattering, the 
S (sideways) direction in the plane of scattering and normal to the beam,  
and the L direction (Longitudinal) along the beam. 
These protons scattered from a longitudinally polarised target \cite {Shypit}. 
The data determined magnitudes and phases of all partial waves up to 
$^3F_3$ \cite {Hasan}, except that  the phase of one small partial wave, 
$^3F_2$, needed to be constrained to a calculation done  by
Blankleider \cite {Blankleider}, and extended later by Blankleider and
Afnan \cite {Afnan}. 
Earlier than this, Hoenig and Rinat predicted in 1974 that loops would 
appear on the Argand diagrams for $\pi d \to NN$ amplitudes from 
projection of the $\pi N$ $P_{33}$ resonance when $\pi d$ kinematics 
and Fermi motion are folded in \cite {Rinat}; that was precisely how it
turned out eventually.
   
An extension of the LAMPF experiment determined the spin dependence 
of $pp \to np\pi^+$ from 492 to 796 MeV lab kinetic energy \cite {Silbar}.
Spin correlation parameters $A_{LL}, A_{SL}, A_{NL}, A_{NO}, A_{SO}$,
$A_{L0}$ and $A_{0L}$ were measured.
The parameter $A_{N0}$ alone is not sufficient to determine the full
Argand plot of amplitudes and their interferences. 
In the representations popularly used, it measures the imaginary part of
the interferences between amplitudes. 
The parameter $A_{S0}$ measures the real part of the same interferences.
There is a general theorem for all final states that $A_{SL}$ measures
the imaginary part of exactly the same amplitudes and is therefore a very
powerful constraint. 
There is limited sensitivity in the parameter $A_{NL}$ because of the
longitudinal orientation of the target polarisation. 

Results from this experiment ruled out broad dibaryons in this mass range
in the dominant $NN$ $^3P_2$ and $^3F_3$ partial waves and smaller
partial waves were negligibly small \cite {Shypit2}.
In the $^1D_2$ amplitude, a phase variation of $33^\circ$ was observed.
The conclusion was that the $360^\circ $ phase variation required for a
dibaryon resonance was definitely ruled out. 
A further remark is that there is presently no evidence for the existence of
the $H$ dibaryon.

It would be very important if $\Delta \Delta$ dibaryons exist.
However, this needs to be proved. 
The WASA data presently contain only one phase sensitive measurement, 
$A_{N0}$. 
This is inadequate for a full determination of all partial waves. 
The key point is that a claim for the existence of a dibaryon resonance
must map out the phase variation  in the Argand diagram and prove the
existence of a pole. 
There are data for $A_{xx}$ and $A_{yy}$, but these depend only on
intensities.
So there is insufficient data at present to map out Argand diagrams
for individual partial waves.
Without these Argand diagrams, there is no proof of the existence of
a dibaryon. 
What is needed is to use a polarised proton beam to study the same
set of parameters as were measured at LAMPF. 
It would require the use of a solenoid to rotate the beam polarisation and a
polarised target. 
This  would extend enormously the physics content of the data. 
From such an experiment it should be possible to establish in a single
run whether the $NN^*(1440)$ channel accounts for the present WASA
data and how it interferes with $\Delta \Delta$ final states.
This should hopefully complete the WASA experiments on this topic.

\end  {document}